\documentclass[prl, twocolumn, aps]{revtex4}

\usepackage{amsmath}

\newcommand{\Lie}[0]{{\cal L}\, }

\newcommand{\be}{\begin{equation}}
\newcommand{\ee}{\end{equation}}
\newcommand{\bea}{\begin{eqnarray}}
\newcommand{\eea}{\end{eqnarray}}

\begin{document}

\title{The First Law for Slowly Evolving Horizons}
\author {Ivan Booth}
\email{ibooth@math.mun.ca}
\affiliation{Department of Mathematics and Statistics, Memorial
 University of Newfoundland \\  
St. John's, Newfoundland and Labrador, A1C 5S7, Canada}
\author{Stephen Fairhurst}
\email{sfairhur@gravity.phys.uwm.edu}
\affiliation{Theoretical Physics Institute, Department of Physics \\
University of Alberta, Edmonton, Alberta, T6G 2J1, Canada}

\begin{abstract}
We study the mechanics of Hayward's trapping horizons, taking isolated
horizons as equilibrium states. Zeroth and second laws of dynamic
horizon mechanics come from the isolated and trapping horizon
formalisms respectively. We derive a dynamical first law by
introducing a new perturbative formulation for dynamic horizons in
which ``slowly evolving" trapping horizons may be viewed as
perturbatively non-isolated.
\end{abstract}

\maketitle

The laws of black hole mechanics are one of the most remarkable
results to emerge from classical general relativity.  Recently they
have been generalized to locally defined isolated horizons \cite{ih}.
However, since no matter or radiation can cross an isolated horizon,
the first and second laws cannot be treated in full generality.
Instead, the first law arises as a relation on the phase space of
isolated horizons rather than as a truly dynamical relationship.  Even
existing physical process versions of the first law \cite{wg, abf}
consider transitions between infinitesimally separated isolated (or
Killing) horizons.  In this paper, we introduce a framework which
allows us to extend the first law to all slowly evolving horizons,
even those whose initial and final states are not infinitesimally
separated.  Additionally, we provide a simple characterization of how
close a horizon is to equilibrium, which will be useful in the final
stages of numerical simulations of black hole collisions.  To obtain
such a law, we first need a local dynamical definition of a black hole
horizon. Hayward has introduced future outer trapping horizons (FOTHs) for
exactly this purpose.  Furthermore, he has shown that they necessarily 
satisfy the second law --- their area cannot decrease in time \cite{hayward}.

In this paper, we will show that both the zeroth and first laws are
also applicable to FOTHs.  In order to do so, we introduce dynamical
notions of surface gravity and angular momentum which are applicable
to all such horizons. It follows immediately that the surface gravity
is necessarily constant if the horizon is in equilibrium
(i.e. isolated).  Next, we introduce the notion of a slowly evolving
horizon, for which the gravitational and matter fields are slowly
changing.  It is only in this limited context that we expect to obtain
the dynamical first law.  We will show that this is indeed the case.

Let us begin by recalling Hayward's definition:

\noindent \textbf{Future Outer Trapping Horizon:} A future outer
trapping horizon (FOTH) is a smooth three-dimensional sub-manifold $H$
of space-time which is foliated by a preferred family of space-like
two-sphere cross-sections $H_v$, with future-directed null normals
$\ell$ and $n$.  The expansion $\theta_{(\ell)}$ of the null normal
$\ell$ vanishes.  Further, both the expansion $\theta_{(n)}$ of $n$
and $\Lie_{n} \theta_{(\ell)}$ are negative.

This definition captures the local conditions by which we would hope
to distinguish a black hole.  Specifically, the expansion of the
outgoing light rays is zero on the horizon, positive outside and
negative inside.  Additionally, the ingoing null rays are converging.
As we will see, these horizons can be space-like or null and include
both equilibrium and non-equilibrium states --- an important feature
for our perturbative study. By comparison, if one is interested only
in the dynamical phase, the space-like dynamical horizons recently
introduced by Ashtekar and Krishnan \cite{ak} are more
relevant. However, such horizons are always expanding and so not so
suitable for studying transitions to and from equilibrium.

Hayward \cite{hayward} has extensively studied the properties of
FOTHs.  Here, we summarize only those of his results which are
important for our work.  First, consider the quantities associated
with the null vector fields $\ell$ and $n$, which we normalize so that
$\ell \cdot n = -1$.  We denote their relative expansions
$\theta_{(\ell)}$ and $\theta_{(n)}$. Their twists are zero since they
are normal to the $H_v$ cross-sections of the horizon. Finally, we
write their shears as $\sigma^{(\ell)}_{ab}$ and $\sigma^{(n)}_{ab}$.

Next, for each choice of the fields $\ell$ and $n$, there exists a
scalar field $C$ on $H$ so that
\begin{equation}\label{v}
   \mathcal{V}^{a} = \ell^{a} - C n^{a} \quad \mbox{and} \quad
   \tau_{a} = \ell_{a} + C n_{a} \, , 
\end{equation}
are respectively tangent and normal to the horizon. Note that
$\mathcal{V} \cdot \mathcal{V} = - \tau \cdot \tau = 2C$. Hayward
\cite{hayward} has shown that if the null energy condition holds, then
\begin{equation}\label{cge0}
   C \ge 0
\end{equation}
on a FOTH.  Thus, the horizon must be either space-like or null, and
the second law of trapping horizon mechanics follows quite easily. If
$\tilde{q}_{ab}$ is the two-metric on the cross-sections, and
$\sqrt{\tilde{q}}$ is the corresponding area element, then
\begin{equation}\label{metricdetderiv}
   \Lie_{\mathcal{V}} \sqrt{\tilde{q}} = - C \theta_{(n)}
   \sqrt{\tilde{q}} \, .  
\end{equation}
By definition $\theta_{(n)}$ is negative and we have just seen that
$C$ is non-negative.  Hence we obtain the local second law: If the
null energy condition holds, then the area element $\sqrt{\tilde{q}}$
of a FOTH is non-decreasing along the horizon \cite{hayward}. Clearly
on integration, the same law applies to the total area of the
horizon. It is non-decreasing and remains constant if and only if the
horizon is null.

To further restrict the rescaling freedom of the null vectors, we
require that $\Lie_{\mathcal{V}} v = 1$, where $v$ is the foliation
parameter labeling the cross-sections. Thus, choosing the foliation
parameter fixes the length of $\mathcal{V}^a$, $\tau^a$, and the null
vectors.  However, at this stage we are still free to change the
foliation labeling and, by doing so, rescale $\mathcal{V}^a$ and the
null vectors by functions of $v$.  From the isolated horizon
perspective, the normalization we have chosen is very natural, as in
that case $\mathcal{V}^{a} = \ell^{a}$ and it is customary to choose
the foliation parameter $v$, so that $n = - dv$.

\noindent\textit{Physical characterization of trapping horizons} ---
The laws of black hole mechanics are given in terms of physical
quantities such as energy, angular momentum, surface gravity and
area. One of the advances of the isolated horizon formalism was to
provide definitions of all these quantities at the horizon, without
reference to space-like infinity or the space-time in which the
horizon is embedded \cite{abf, ih}. In generalizing these definitions
to FOTHs, we will be motivated by two requirements: i) the new
definitions should match the old ones when trapping horizons are
isolated and ii) the new definitions should depend only on
quantities that are intrinsic to the horizon, and as such truly be
properties of the horizon itself. That said, the ultimate
justification for these expressions will be found in their utility in
the calculations that follow.

For isolated horizons both surface gravity and angular momentum were
defined with respect to a one-form $\omega_{a}$ which can be written
as:
\begin{equation}
\label{omega}
  \omega_{a} := - n_{b} \underleftarrow{\nabla}_{a} \ell^{b} \, ,
\end{equation}
where the arrow signifies that the derivative is pulled back to the 
horizon.  Then, the isolated horizon surface gravity is given by
$\kappa_{\ell} = \ell^{a} \omega_{a}$ while angular momentum
information is contained in the other components of $\omega_{a}$.

Now, $\omega_a$ written in this form (equation (\ref{omega}))
continues to be an intrinsically well-defined quantity for trapping
horizons. Then, an obvious generalization is to define:
\begin{eqnarray}\label{angmom}
\kappa_{v} &:=& \mathcal{V}^a \, \omega_a \,\,(\,= -n_b\,
\mathcal{V}^{a} \,\nabla_{a} \, \mathcal{V}^{b}\,) \quad \mbox{and}
\nonumber \\ 
J_\varphi &:=& \frac{1}{8 \pi} \int_{H_v} d^2 x \sqrt{\tilde{q}}\,
\varphi^a \,\tilde{\omega}_a  \, ,
\end{eqnarray}
where $\varphi^{a}$ is any vector field tangent to the horizon
cross-sections, and $\tilde{\omega}_{a}$ is the projection of
$\omega_{a}$ into the 2-surface $H_{v}$.  In the case where
$\varphi^{a}$ is divergence free, $J_\varphi$ is independent of the
normalization of $\ell$ and equal to other popular expressions for angular
momentum (such as the Komar, Brown--York \cite{by} or
Ashtekar--Krishnan \cite{ak} definitions).  It is clear that both the
surface gravity $\kappa_{v}$ and angular momentum $J_{\varphi}$
expressions reduce to their isolated horizon values if the horizon is
null.

\noindent\textit{The constraint law} --- Consider the set of vector
fields in $H$ that generate one-parameter families of diffeomorphisms
that map two-dimensional cross-sections $H_{v}$ of the horizon into
each other. Any such vector field may be written in the form $ X^a =
x_{o} \mathcal{V}^a + \tilde{x}^a $, for some function $x_{o}(v)$ and
vector field $\tilde{x}^a$ that is everywhere tangent to an
appropriate cross-section.  Then, by integrating the $\tau^{a} X^{b}$
component of the Einstein equations over $H_{v}$, we obtain the
following relationship:
\begin{eqnarray}\label{constraint}
    &&\frac{1}{8 \pi G} \int_{H_v} d^2 x \left[ \kappa_{v}
      \Lie_X \sqrt{\tilde{q}} + \tilde{x}^{a} \Lie_{\mathcal{V}} \, 
      (\sqrt{\tilde{q}}\tilde{\omega}_{a}) \right] \\
   &&\quad = \int_{H_v} d^2 x \sqrt{\tilde{q}} \left[T_{ab} X^a
      \tau^b \right] \nonumber \\
   &&\quad + \frac{1}{16 \pi G} \int_{H_v} d^2 x
      \sqrt{\tilde{q}}(\sigma^{(\ell)ab} 
      + C \sigma^{(n)ab}) (\Lie_{X} \tilde{q}_{ab})\nonumber \\   
   &&\quad\!\!\! + \frac{1}{16 \pi G} \int_{H_v} d^2 x \left[ 2C
      \sqrt{\tilde{q}} (\Lie_{X} \theta_{(n)}) + C 
      \theta_{(n)} (\Lie_{X} \sqrt{\tilde{q}}) \right] .
      \nonumber 
\eea
A full derivation of this result will be given in \cite{bfbig} and in
another paper we will see that this relation plays a crucial role in
the Hamiltonian formulation of general relativity on manifolds with
FOTHs as boundaries \cite{bfham}. Here, we show that in certain
restricted situations equation (\ref{constraint}) becomes a first law
for dynamical black holes.

\noindent\textit{Quasi-stationary horizons} --- At first glance, one
might think that (\ref{constraint}) is already the dynamical first law
of black hole mechanics.  After all, it relates rates of change of
area and angular momentum to fluxes across the horizon.  However,
there are several reasons why this is not so.  The first and most
important is that we should not expect the standard $\kappa \dot{a}$
form of the first law to hold in all dynamical situations.  In
thermodynamics, it is only in the quasi-static case that it is
possible to write the energy balance equation as
\begin{equation}\label{thermofirstlaw}
   dE = TdS + \mbox{work terms} \, .
\end{equation}
Furthermore, in the general case, there is no clear interpretation of
the right hand side of (\ref{constraint}) as a flux of energy through
the horizon. Instead, we will require the horizon to be
``quasi-stationary'' and then obtain a first law. 
 
Heuristically, it is clear that the properties of a quasi-stationary
horizon, such as the area and surface gravity , should be slowly
varying.  However, since there is a rescaling freedom in the vector
field $\mathcal{V}$, requiring $(\Lie_{\mathcal{V}}
\sqrt{\tilde{q}})/\sqrt{\tilde{q}}$ to be small is not a meaningful
condition --- it can be satisfied on \textit{any} trapping horizon by
suitably rescaling $\mathcal{V}$.  Furthermore, we cannot fix the norm
of $\mathcal{V}$ (or its average) to unity since we are interested in
the limit as $\mathcal{V}$ becomes null.  Instead, we introduce the
one-form $\chi_{a}:= d_{a}v$ satisfying $\mathcal{V} \cdot \chi = 1$,
and require $(\Lie_{\mathcal{V}} \sqrt{\tilde{q}}) (\Lie_{\chi}
\sqrt{\tilde{q}})/\tilde{q}$ to be small compared to the
characteristic scale of the horizon.  This condition is invariant
under rescalings and the expression vanishes whenever the horizon
is null. On a space-like horizon, $(\Lie_{\mathcal{V}}
\sqrt{\tilde{q}}) (\Lie_{\chi} \sqrt{\tilde{q}}) = \tilde{q} \,C
\,\theta_{(n)}^2$ and so is closely related to the expansion of the
surface (see equation (\ref{metricdetderiv})).  By evaluating this
expression, we can invariantly determine the expansion rate of the
horizon and choose the foliation accordingly.  Then, we require other
fields to be slowly varying with respect to this foliation.

\noindent \textbf{Slowly Evolving Horizon:} A region of a future outer
trapping horizon $H$ with $v \in [v_{1},v_{2}]$, is slowly evolving
(at a rate $\epsilon$) if there exists a parameter $\epsilon \ll 1$
such that,

\begin{enumerate}

\item On every cross section $H_v$ with $v \in [v_{1}, v_{2}]$,\\
  $\displaystyle
  \int_{H_{v}} d^{2}x \sqrt{\tilde{q}} \left(\tilde{q}^{ab}\,
  \nabla_{a}\, \mathcal{V}_{b} \right) \left(\tilde{q}^{cd}\,
  \nabla_{c}\, \mathcal{\chi}_{d} \right) \le \epsilon^{2} \, .$
\item The foliation parameter $v$ is chosen so that $|\mathcal{V}|
  = \sqrt{2C} \sim \epsilon$, and its rates of change are similarly
  small over the horizon.

\item The one-form $\tilde{\omega}_{a}$ and expansion $\theta_{n}$ are
  slowly evolving: $\displaystyle |\Lie_{\mathcal{V}} \,
  \tilde{\omega}_a| \le \epsilon/r_{H}^{2}$ and $\displaystyle
  |\Lie_{\mathcal{V}} \, \theta_{(n)}| \le \epsilon/r_{H}^{2}$, where
  $r_{H}$ is the area radius of the horizon, $a_{H} = 4\pi r_{H}^{2}$. 

\item $\displaystyle{|\tilde{\mathcal{R}}|, \, |\omega|^2, \,
  |\sigma^{(n)}|^2}$ and $T_{ab} n^a n^b \sim 1/r_{H}^{2}$ or smaller.

\end{enumerate}

The first condition gives an invariant characterization of slow
expansion. The second condition restricts the foliation to guarantee
that the area is slowly evolving with respect to $\mathcal{V}^{a}$,
specifically $(\Lie_\mathcal{V} \, r_{H}) \sim \epsilon^{2}$.  The
third requires other geometric quantities to also be slowly evolving,
while the fourth fixes a reasonable horizon geometry and demands that
conditions in the surrounding space-time not be too extreme, here
$\tilde{\mathcal{R}}$ is the Ricci scalar of the two-surfaces $H_{v}$.
An immediate consequence of our definition is that isolated horizons
are examples of slowly evolving horizons with $\epsilon = C = 0$
(provided condition 4 is satisfied).

Let us now consider the implications of this definition. Here, we will
simply state the consequences, a more complete discussion will be
given in \cite{bfbig}.  For concreteness, we restrict the allowed
matter to be scalar and/or electromagnetic fields. From a projection
of the Einstein equations and condition 2, we can show that
\begin{equation}\label{shearandtll}
   |\sigma^{(\ell)}_{ab}| \sim \epsilon
\quad \mbox{and} \quad 
 T_{ab} \ell^{a} \ell^{b} \sim \epsilon^{2}.
\end{equation}
(Here and in the following, we choose to focus on the
$\epsilon$ factors. The powers of $r_{H}$ required to make the equations 
dimensionally correct are omitted.)  Then, the allowed
form of the matter fields forces $T_{ab} \ell^{a} \tilde{x}^{b} \sim
\epsilon$ and $\Lie_{\mathcal{V}} (T_{ab} \ell^{a} n^{b}) \sim
\epsilon$.  Further application of the Einstein equations and Bianchi
identities gives the Weyl components $\Psi_{0}, \Psi_1 \sim \epsilon$
and $\Lie_{\mathcal{V}} \tilde{\mathcal{R}} \sim \epsilon$. On an
isolated horizon each of these quantities would be 
zero.  Lastly, condition 3 is sufficient to guarantee that
\begin{equation}\label{kappaderiv}
   |\Lie_{\mathcal{V}} \, \kappa_{v}| \sim \epsilon
   \quad \mbox{and} \quad 
   |\tilde{q}_{a}^{\,\, b} \nabla_{b} \, \kappa_{v}| \sim \epsilon \, .
\end{equation}
Therefore, on a slowly evolving horizon, the surface gravity is a
slowly varying function and we can write
\begin{equation}
   \kappa_{v} = \kappa^{(0)} + \epsilon \kappa^{(1)} \, ,
\end{equation}
where $\kappa^{(0)}$ is a constant.  Recall that an isolated horizon
corresponds to $\epsilon = 0$.  Hence, we immediately obtain the
zeroth law: the surface gravity of a FOTH is constant if
the horizon is isolated. 

\noindent\textit{The dynamical first law} --- We can now derive the
first law for slowly evolving horizons. First we consider evolution
with respect to $\mathcal{V}^a$, an appropriate evolution vector for a
horizon with no obvious axis of symmetry --- a non-rotating horizon.
Setting $\tilde{x}^a = 0$ and $x_{o} = 1$ in the constraint law
(\ref{constraint}) and expanding in powers of $\epsilon$, we find that
all terms vanish at zeroth and first order.  Then, to
$\mathcal{O}(\epsilon^{2})$, the first law for a slowly evolving
horizon reads:
\begin{equation}\label{firstlaw}
   \frac{ 1}{8\pi G} \kappa^{(0)} \dot{a}_H = \dot{E} :=
   \int_{H_v} d^{2}x \sqrt{\tilde{q}} \left[ T_{ab} \ell^{a} 
   \ell^{b} + \frac{1}{8\pi G} |\sigma^{(\ell)}|^2  \right] \, ,
\end{equation}
where the dot signifies a derivative with respect to
$v$.  It says that the surface gravity multiplied by the
change in area is equal to a flux of energy through the horizon.  This
flux is comprised of two terms, both of which are positive.  The first
is the flux of matter through the horizon and is familiar from
standard physical process versions of the first law \cite{wg}.  The
second term is a flux of gravitational shear through the horizon ---
which would naturally be interpreted as a flux of gravitational
radiation.  A similar term has been obtained previously by Hawking
\cite{hawking} when considering perturbations of an event horizon.

We would like to ``integrate'' (\ref{firstlaw}) in order to obtain an
expression for the energy of the horizon.  To this end, we further
restrict the choice of $\mathcal{V}^{a}$ by requiring that:
\begin{equation} 
   \kappa^{(0)} = \frac{1}{2 r_{H}} \, .
\end{equation}
Note that while we cannot prove that this rescaling is possible on
every slowly evolving horizon, we can show that if the horizon
satisfies a certain genericity condition, it will be. In particular,
perturbations to Schwarzschild and non-extremal Kerr horizons will
satisfy the condition.  Since $\kappa^{(0)}$ is a function of $r_{H}$
alone, we can integrate the $\kappa \,\dot{a}$ term in (\ref{firstlaw})
to obtain the usual expression for the energy
\begin{equation}\label{energy}
   E = \frac{r_{H}}{2G}  \, .
\end{equation}

\noindent\textit{Rotating horizons} --- Next, consider rotating
horizons.  In (\ref{angmom}) we gave a definition for the angular
momentum.  However, this quantity only has a physical significance if
the vector field $\varphi^{a}$ is an (approximate) symmetry of the
horizon.  Hence, we define an

\noindent \textbf{Approximately symmetric horizon:} The vector field
$\varphi^{a}$ is an approximate symmetry of a trapping horizon if

\begin{enumerate}

\item $\varphi^a$ is tangent to the cross sections of the horizon,
generates a family of closed integral curves and is normalized so that
those curves have affine parameter length $2 \pi$.
\item $\varphi^{a}$ is Lie dragged up the horizon by
  $\mathcal{V}^{a}$, i.e. $\Lie_{\mathcal{V}} \, \varphi^{a} = 0$.
\item $\varphi^{a}$ is an approximate symmetry of the
  horizon geometry and matter fields:
$|\tilde{q}^{ab} \nabla_{a} \varphi_{b}| < \epsilon^{2}$, 
$   |\Lie_{\varphi} \,\tilde{q}_{ab}| < \epsilon$, 
$   |\Lie_{\varphi} \,\tilde{\omega}_{a}| <\epsilon$, 
$   |\Lie_{\varphi} \,\theta_{(n)}| < \epsilon$ and 
$   |T_{ab} \varphi^{a} \ell^{b}| < \epsilon^{2}
   \, .$   
(These conditions are analogous to those satisfied by
$\mathcal{V}^{a}$ on a slowly evolving horizon).
\end{enumerate}

For a slowly evolving horizon which is approximately symmetric, the
angular momentum $J_{\varphi}$ is a meaningful quantity.  Going back
to the constraint law and setting $x_{o}=0$ and $\tilde{x}^{a} =
\varphi^{a}$ we find its time rate of change.  As in the non-rotating
case, equation (\ref{constraint}) will vanish at zeroth and first
orders in $\epsilon$.  To second order we have:
\begin{equation}\label{angmomchange}
   \Lie_{\mathcal{V}} \, J_{\varphi} = \int_{H_v} d^{2}x
   \sqrt{\tilde{q}} \left[ T_{ab} \tau^{a} \varphi^{b} +
   \frac{1}{16 \pi G} \sigma^{(\ell)ab} (\Lie_{\varphi} \tilde{q}_{ab}) \right] \, .
\end{equation}
Thus, the rate of change of angular momentum of the horizon depends
upon the flux of matter and gravitational fields through the horizon.

To get a general first law, we would like to combine the rate of
change of angular momentum with the first law for non-rotating
horizons (\ref{firstlaw}).  To do this, we again fix the average value
of the surface gravity as well as an angular velocity $\Omega$ (we are
following a similar strategy to that taken in \cite{ak}, though with
the caveat that here only the average value of the surface gravity is
fixed).  We do this by requiring that both take the same values as in
a Kerr space-time with the same $J_{\varphi}$ and $a_{H}$:
\begin{equation}\label{intkapparot}
   \Omega := \frac{2 G J_{\varphi}}{r_{H} \sqrt{r_{H}^{4} + 4 G^{2}
   J_{\varphi}^{2}}} \quad \!\! \mbox{and} \!\!\quad  
   \kappa^{(0)} = \frac{r_{H}^{4} - 4 G^{2} J_{\varphi}^{2}}{2
   r_{H}^{3} \sqrt{r_{H}^{4} + 4 G^{2} J_{\varphi}^{2}}}  .
\end{equation}
With area and angular momentum slowly changing up the horizon, the
angular velocity is also slowly varying.

Finally, we consider the full constraint law (\ref{constraint}) with 
the evolution vector field $X^{a}$ chosen as
\begin{equation}
   t^{a} := \mathcal{V}^{a} + \Omega \varphi^{a} \, .
\end{equation}
Then, we can once again expand out the constraint law
(\ref{constraint}) to order $\epsilon^{2}$ and this time obtain:
\begin{eqnarray}\label{physproc1law}
     \dot{E} &=& \frac{1}{8 \pi G} \kappa^{(0)} \dot{a} + \Omega
     \dot{J}_\varphi \quad \mbox{where}  \\ 
     \dot{E} &:=& \int_{H_v} d^{2}x \sqrt{\tilde{q}} \left[ T_{ab} t^{a}
     \tau^{b} + \frac{1}{16 \pi G}  \sigma^{(\ell) ab} \left(\Lie_{t} \tilde{q}_{ab}
     \right)\right] \, .\nonumber
\end{eqnarray}
Since $\kappa^{(0)}$ and $\Omega$ are specific functions of only
$r_{H}$ and $J_{\varphi}$, it is once again possible to ``integrate''
the left hand side of the equation to obtain as an expression for the
energy
\begin{equation}
   E = \frac{\sqrt{r_{H}^{4} + 4 G^{2} J_{\varphi}^{2}}}{2 G
   r_{H}} \, .
\end{equation}
This $E$ is equal to the energy of a rotating isolated horizon with
the preferred choice of normalization for $\ell$, and in particular is
the mass of a Kerr black hole with parameters $a_{H}$ and
$J_{\varphi}$.

The notion of a slowly evolving horizon provides a bridge between the
equilibrium of isolated horizons and the fully dynamical horizons of
\cite{ak}.  In particular, our first law (\ref{physproc1law}) bears a
striking resemblance to the dynamical horizon energy balance
formula. There are, however, several important differences.  Firstly,
we restrict to slowly evolving horizons, which are near to
equilibrium.  Doing so enables us to provide a locally defined surface
gravity $\kappa_{v}$ which is shown to be slowly varying over the
horizon.  This surface gravity maintains its usual interpretation as
the acceleration of a vector field, $\mathcal{V}^{a}$, along the
horizon.  Secondly, we do not obtain an equivalent of the
$|\zeta|^{2}$ term of \cite{ak}, which is likely related to angular
momentum.  However, we expect that term would vanish at the order of
perturbation theory which we are considering. In the future, we plan
to examine the connection between these two formulations in more
detail and determine in what precise sense our first law can be seen
as a perturbative form of the dynamical horizon energy balance
formula.

To summarize, we have examined conditions for which a slowly evolving
FOTH may be said to be ``perturbatively non-isolated". This has
allowed us to obtain a truly dynamical version of the first law of
black hole mechanics.  However, we expect that the notion of a slowly
evolving horizon will also find application in numerical studies of
how horizons settle down to equilibrium.  Specifically, one could
track the approach of the parameter $\epsilon$ to zero.

We would like to thank A.~Ashtekar, C.~Beetle and B.~Krishnan for
discussions.  SF was supported by the Killam Trusts at the University
of Alberta and NSF grant PHY--0200852 at UWM; IB was supported by NSERC.

\end{document}